# Assembly of High-Performance van der Waals Devices Using Commercial Polyvinyl Chloride Films

Son T. Le,[1,2,*,#] Jeffrey J. Schwartz,[1,2,3,*,#] Tsegereda K. Esatu,[1,2] Sharadh Jois,[2] Andrea Centrone,[3] Karen E. Grutter,[2] Aubrey T. Hanbicki,[2] Adam L. Friedman[2,*]

[1]Department of Electrical and Computer Engineering, University of Maryland, College Park, Maryland 20742, United States

[2]Laboratory for Physical Sciences, College Park, Maryland 20740, United States

[3]Physical Measurement Laboratory, National Institute of Standards and Technology, Gaithersburg, Maryland 20899, United States

[#]S.T.L. and J.J.S. contributed equally

[*]Corresponding author emails:
stle@lps.umd.edu, jjschwa@lps.umd.edu, afriedman@lps.umd.edu

**Abstract**

Control over the position, orientation, and stacking order of two-dimensional (2D) materials within van der Waals heterostructures is crucial for applications in electronics, spintronics, optics, and sensing. The most popular strategy for assembling 2D materials uses purpose-built stamps with working surfaces made from one of several different polymers. However, these stamps typically require tedious preparation steps and suffer from poor durability, contamination, and limited applicability to specific 2D materials or surfaces. Here, we demonstrate significant improvements upon current 2D flake transfer and assembly practices by using mechanically durable stamps made from polyvinyl chloride (PVC) thin films. These stamps are simpler to prepare compared with existing methods and can withstand multiple transfer cycles, enabling greater reusability. We use two commercially available PVC films with distinct pick-up and release temperatures. Together, these films also enable polymer-to-polymer flake transfers and stack-and-flip fabrication of inverted heterostructures in one seamless process. Systematic comparisons of cleaning processes confirm the removal of PVC-derived residue from the assembled structures to create atomically clean interfaces. We demonstrate the utility and versatility of these polymer films and transfer process by fabricating graphene/hexagonal boron nitride heterostructure devices with high-performance electrical characteristics. Further, we demonstrate the ability to pick up and to deposit bulk aluminum gallium arsenide nanostructured films, enabling the creation of heterogeneously integrated devices. These advances enable faster and more versatile assembly of 2D materials than previously reported polymer-assisted methods. Accordingly, this technique increases fabrication rates, improves device quality, and enables more complex structures, thereby facilitating nanomaterial assembly in a broad range of applications.

The isolation of two-dimensional (2D), van der Waals (vdW) materials[1–3] kickstarted a new era of materials and device research.[4] These materials, which have strong intralayer bonding but weak interlayer coupling, can be exfoliated into flakes as thin as a single atomic layer. After exfoliation, flakes of different materials can be discretely stacked to create vdW heterostructures with properties and compositions that are not achievable through conventional growth methods.[5,6] In this way, researchers can create novel materials and structures with precisely engineered characteristics that are useful in a wide range of electronics, spintronics, optics, and sensing applications.[7–9]

Common strategies[10,11] to assemble vdW heterostructures, include both wet[12] and dry[13] polymer-assisted techniques. These techniques leverage the temperature-dependent adhesive properties of polymer surfaces to pick up, stack, and deposit exfoliated flakes. Ongoing advances in all-dry, polymer-assisted assembly[14] and device fabrication[15] enable increasingly complex vdW heterostructures and high-quality integrated devices.[6,13,16–18]

The principle behind polymer-assisted assembly of 2D materials is elegantly simple. An elastic stamp with a polymer working surface is brought into direct contact with a flake. The adhesion strength between the flake and the polymer stamp depends, in part, on the temperature of this junction. Accordingly, controlling the temperature of its working surface enables a stamp to pick up and to deposit 2D flakes on demand. Furthermore, the same stamp can pick up flakes in series to assemble vdW heterostructures with controlled sequences, positions, and orientations. Depositing these precisely assembled structures onto suitable substrates, and leveraging other nanofabrication techniques, enables their integration into functional devices.

The mechanical procedure of using polymer stamps to assemble vdW heterostructures has not changed significantly since it was first implemented more than 10 years ago.[13] Nevertheless, significant effort has been spent to discover improved polymer surfaces for stacking 2D materials.[19] One of the more popular polymers employed for this purpose is polypropylene carbonate (PPC)[13] since it performs particularly well at picking up and releasing hexagonal boron nitride (hBN), a commonly used 2D insulator. Propylene carbonate (PC),[20] polymethyl methacrylate (PMMA),[21] and even nail polish[22] also function as working polymers.[19] A significant drawback of these materials, however, is the tedious, time-consuming preparation of the thin films required for each stamp, which are mechanically fragile and easily damaged. Repeated deformations of the polymer during flake stacking and transfer strain the film, which can tear or break, ruining the stamp and any structures being assembled. Furthermore, these polymers only work well to pick up a few types of 2D materials, while other materials are not reliably transferred, thereby limiting the compositions of assembled heterostructures. For this reason, hBN, which is typically easy to pick up, is sometimes used to transfer other 2D materials (e.g., graphene) that are less reliably picked up with polymer surfaces. This practice constrains vdW heterostructure design by placing an insulating (hBN) layer at the top of the stack. Additionally, the optimal temperatures for 2D material pick up (from the substrate to the stamp) and release (from the stamp to the substrate) are typically fixed for each polymer, thereby reducing flexibility in operating conditions during the fabrication process.

Recently, polyvinyl chloride (PVC) has emerged as an promising polymer for 2D material transfer and vdW heterostructure assembly.[23–25] Unlike PPC, which typically picks up and releases 2D materials at fixed temperatures, PVC was shown to be able to pick up and release hBN at working temperatures that strongly depend on the film thickness. In a recent report, Onodera and coworkers found that thinner PVC films pick up and release hBN at higher pick-up and release temperatures than thicker PVC films.[23] Using this observation, the authors demonstrated all-dry polymer-to-polymer (P2P) transfer using specially formulated and cast PVC films to assemble hBN flakes. This new capability has the potential to foster numerous applications and to enable measurements of vdW heterostructures that are prohibitively challenging to assemble using other working polymers.[17,26]

Despite the significant potential of PVC-based stamps for versatile 2D material assembly, early demonstrations of the technique required tedious, time-consuming preparation steps such as the custom-formulation and casting of polymer films. These steps greatly discourage broader use of PVC stamps due to the complicated process controls required to achieve consistent film characteristics suitable for transferring 2D flakes. A commercial PVC film[24] that was used for this process suffers, in practice, from inconsistent pick-up and deposition characteristics, making P2P transfer difficult and impractical to execute reliably. Additionally, no functional devices were assembled using PVC films in prior works (only hBN/hBN stacks). To demonstrate the utility of PVC-assisted transfer, we fabricate gate-modulate graphene transistors, representative of practical, functional devices commonly studied by the 2D materials community. These devices, assembled using PVC stamps, perform as well as (or better than) devices produced with other common vdW heterostructure assembly methods. Testing practical devices is important since defects and contaminants derived from the assembly process typically adversely affect device performance. We examine polymer residue transferred by the PVC stamps and demonstrate effective cleaning strategies that ensure high-quality, residue-free devices. This detailed study of contaminant transfer and removal, not previously reported for PVC or (to the best of our knowledge) any commonly used polymer for heterostructure assembly, provides important insights for fabricating and cleaning high-quality 2D material devices.

In this manuscript, we demonstrate the use of two commercial PVC films to assemble 2D materials. These films, each mass-produced with consistent characteristics and sold in large rolls, replace manually prepared polymer films that are typically cast in small, custom batches. Further, the two commercial films employed here, which have different and complementary working temperature ranges, enable reliable P2P "stack-and-flip" 2D material assembly, which is otherwise challenging to achieve through other means. Additionally, for the first time, we use PVC stamps to pick up bulk aluminum gallium arsenide nanostructured films and to deposit them precisely on prefabricated a silicon nitride optical resonator. This heterogeneously integrated photonic device exemplifies the outstanding versatility of PVC-assisted assembly, which enables fabrication of hybrid nanostructures that combine different materials technologies in the same assembled device. These demonstrations significantly expand the capabilities of polymer-assisted 2D material and nanostructure assembly. Accordingly, we anticipate widescale impact and broad adoption of these commercial films for rapid fabrication of complex vdW heterostructures and devices.

## Results & Discussion

### PVC Stacking & Transfer

The stamps used here consist of a thin PVC film supported by a convex, dome-shaped scaffold positioned near the end of a transparent cantilever. Compared with flat stamps, domed surfaces enable more precise control over the contact position and area when manipulating 2D flakes, facilitating the assembly of higher quality vdW heterostructures.[27] Additional details of the stamp construction and stacking setup are provided in Methods and in the Supplementary Information (Figures S1 and S2). Importantly, the working surfaces of our stamps are made from one of two different commercially available PVC films, hereafter referred to as "PVC1" and "PVC2" (see Methods). Although the processes we describe could also, potentially, utilize other commercial or custom-made polymer films, we find that the characteristics of these specific PVC films are well-suited for the majority of 2D flake transfer scenarios. Significantly, these films have different working temperature ranges, bracketed by their respective pick-up ($T_{\text{pick up}}$) and release ($T_{\text{release}}$) temperatures. Due to its lower working temperature range ($T_{\text{PVC1 pick up}} \approx 45$ °C and $T_{\text{PVC1 release}} \approx 90$ °C) and its adhesive backing layer, PVC1 is used in most of the 2D stacking tasks. Compared to the commercial film used in previous work,[24] the commercially available PVC1 used here exhibits more favorable pick-up and deposition characteristics (lower temperatures and more reliable transfer). The higher working temperature range of PVC2 ($T_{\text{PVC2 pick up}} \approx 90$ °C and $T_{\text{PVC2 release}} \approx 160$ °C), with a pick-up temperature complementing the release temperature of PVC1, enables P2P transfer and stack-and-flip assembly of 2D materials when pairing stamps made with each polymer.

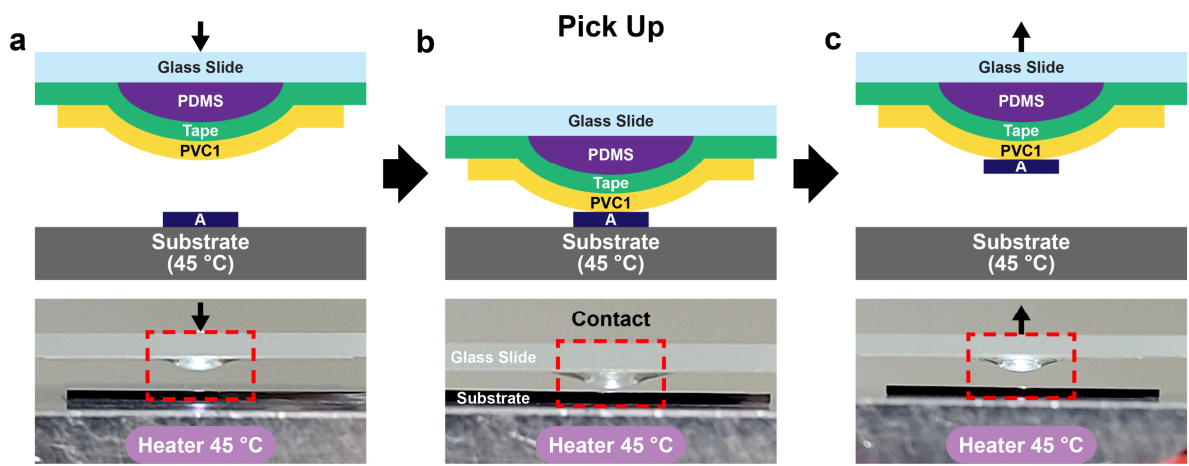

**Figure 1:** Picking up two-dimensional (2D) material flakes using a polyvinyl chloride (PVC) stamp. Here, the stamp consists of a thin PVC1 film supported by a dome-shaped scaffold, made from polydimethylsiloxane (PDMS) and transparent tape, positioned near the end of a glass slide; see also Figure S1. To pick up a 2D flake (material A), the stamp is (a) aligned with the target and then (b) brought into direct contact with the flake while heating the substrate. At a temperature of $\approx 45$ °C, (c) the 2D flake transfers to the stamp when lifted from the substrate. Repeating this process with the same stamp can be used to assemble multilayer van der Waals heterostructures. This process could also be completed with a PVC2 stamp, which has pick-up and release temperatures of $\approx 90$ °C and $\approx 160$ °C, respectively. The regions near the stamp and substrate surfaces, enclosed by dashed red rectangles in the photographs (bottom), are illustrated more clearly in the schematics (top), but are not drawn to scale. The glass slides seen in the photographs are approximately 1 mm thick.

To pick up a 2D flake, first, a stamp is positioned above the target flake (Figure 1a and Figure S3a). Next, the domed working surface of the stamp is brought into direct contact with the target flake (Figure 1b and Figure S3b). At temperature $T_{\text{pick up}}$, the adhesion between the flake and the PVC film is stronger than the adhesion between the flake and underlying substrate.

Consequently, as the stamp is slowly removed from the substrate, the flake transfers to the PVC surface (Figure 1c and Figure S3c). This pick-up process can be repeated using the same stamp to assemble stacks of 2D materials, where the previously transferred flake is used to pick up the next flake in the series. Finally, to deposit the flake(s) or heterostructure(s), the stamp is brought into contact with the destination substrate at temperature $T_{release} > T_{pick\ up}$. At this higher temperature, the adhesion of 2D flakes to the PVC film weakens to less than the adhesion between 2D flakes and the substrate, causing the flakes to remain on the substrate upon removal of the stamp, as illustrated in Figure 2a,b.

When assembling vdW heterostructures, particularly structures consisting of many sequentially stacked 2D flakes, PVC offers a significant advantage over other commonly used (but less-durable) polymers. For example, thin films of PPC and PC require thermal annealing after each pick-up step to relieve stress and to maintain the mechanical integrity of the working surface throughout the assembly process. Consequently, stamps are typically unmounted from the transfer apparatus and baked on a hotplate for ≈ 1 min. These stamps then must be remounted and realigned before picking up the next flake or depositing the assembled stack. Without these disruptive, time-consuming precautions to relieve stress, the polymer films are liable to tear or otherwise deform in ways that can ruin the assembled heterostructures. Even with precautions, however, the stamps still have a finite useful life (typically only a few pick-ups) before their surfaces degrade, which limits the complexity and yield of structures that can be assembled. By contrast, PVC-assisted transfer and assembly can be performed in a continuous sequence (Figure 1) without the need to unmount, bake, and remount the stamps, thereby streamlining the assembly process. This streamlined process enables faster, more reliable, and potentially automated[28] fabrication of complicated heterostructures that are not practical with existing methods.

Considering the wide variety of 2D materials and substrates in use, the precise pick-up and deposition parameters are expected to vary based on the relative adhesion strengths of particular material/substrate combinations. Indeed, picking up some materials (e.g., monolayer graphene on $SiO_2$) using any polymer stamp is challenging or not always achievable under practically useful conditions. That said, PVC stamps reliably pick up hBN flakes from $SiO_2$ surfaces, and hBN is known to be effective at picking up other 2D materials (see Table S1 for a summarized list). For this reason, typically the first material picked up (e.g., material "A" in Figure 1) is hBN. Furthermore, hBN can serve as a uniform, atomically flat, and electrically insulating layer to encapsulate subsequent 2D layers in vdW heterostructures. Such encapsulation protects the other 2D materials in the structure from direct contact with the PVC stamp and helps to reduce the effects of polymer residue contamination (see below). In this way, a variety of 2D materials can be assembled on a stamp, either directly on the PVC film or using an hBN assistance layer, to create a wide range of vdW heterostructures. When using PVC1, we find that 2D flakes (e.g., hBN exfoliated on $SiO_2$) are typically picked up by the stamp between 40 °C and 50 °C ($T_{pick\ up}$) and deposited on substrates between 80 °C and 100 °C ($T_{release}$), as illustrated in Figure 1.

**Polymer-to-Polymer Transfer & Flipping**

In the assembly process described above, the sequence of 2D materials in a vdW heterostructure reflects the order in which the stamp picks up each flake. After deposition, the first flake in the sequence is exposed at the top of the stack, while the last flake in the sequence is buried at the bottom, ultimately in contact with the substrate. In some circumstances, flipping the stack and inverting the sequence of materials in the deposited heterostructure can facilitate subsequent processing or measurements. For example, flipping heterostructures originally assembled with an insulating hBN pick-up assistance layer at the top exposes other layers in the stack for electrical contact deposition. Achieving this flipped-stack assembly is challenging with existing techniques that rely on deconstructing the stamps and removing the extremely delicate films carrying the assembled structures. The unsupported films are then re-deposited (assembly side up) onto a substrate with essentially no control over the final position and orientation of the vdW heterostructure.[29] If successful, the heterostructures sit atop a significantly thicker polymer layer that must be removed, without damaging the structures, before they are available for device integration.[17,26]

Alternatively, as illustrated in Figures 2a,c,d and S4, P2P transfer of a vdW heterostructures between separate polymer stamps can effectively flip the stack and invert the flake sequence when deposited. This P2P transfer is accomplished by first assembling a vdW heterostructure (as shown in Figure 2a) on a stamp made with a relatively low-release-temperature film, such as PVC1 ($T_{\text{PVC1 release}} \approx 90\ °C$). This stamp (with the assembled structure) is then brought into contact with a second (bare) stamp with a pick-up temperature greater than or approximately equal to the release temperature of PVC1, such as PVC2 (i.e., $T_{\text{PVC2 pick up}} \gtrsim T_{\text{PVC1 release}}$). Contacting the two stamps at the PVC2 pick-up temperature ($T_{\text{PVC2 pick up}} \approx 90\ °C$) enables the assembled heterostructures to be released by the first stamp (PVC1) while simultaneously adhering to the second stamp (PVC2), as shown in Figure 2c and Figure S4c. In doing so, the sequence of the assembled flakes relative to the polymer surface is inverted (e.g., from A/B/C to C/B/A). Finally, the flipped heterostructure is deposited (Figure 2d and Figure S4c inset) by contacting a substrate at the PVC2 release temperature ($T_{\text{PVC2 release}} \approx 160\ °C$) and slowly removing the stamp.

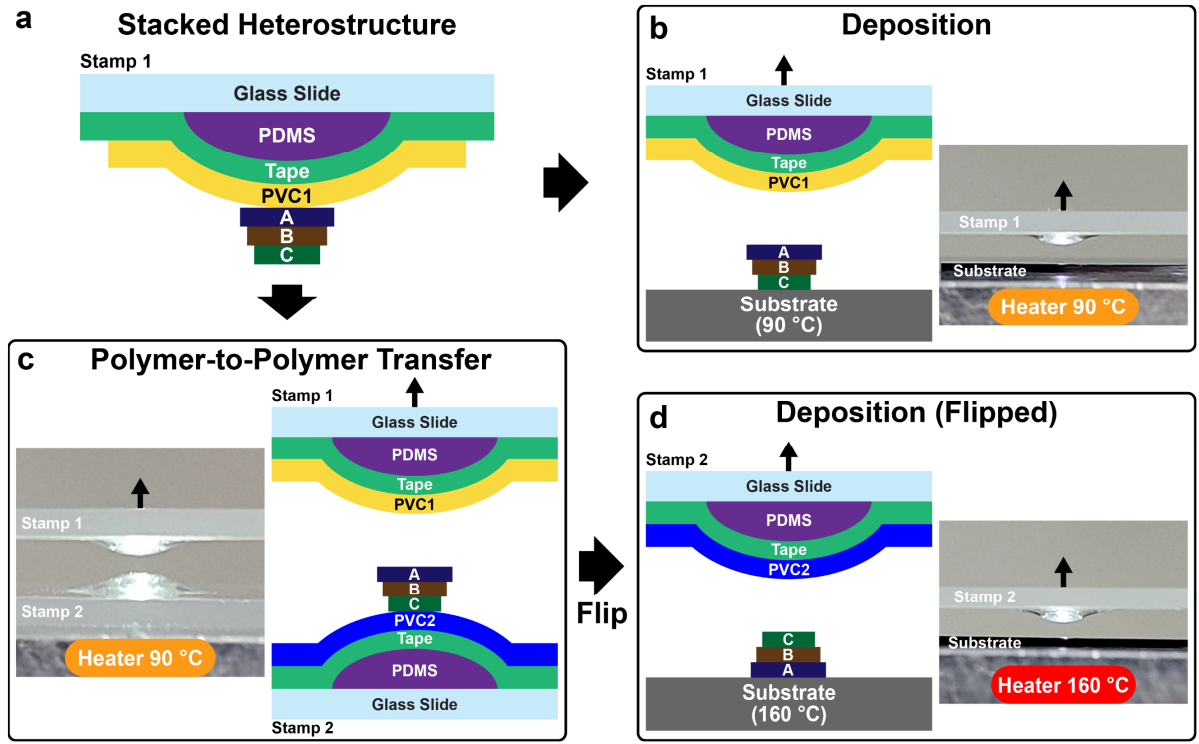

**Figure 2:** Deposition and polymer-to-polymer transfer and flipped-deposition of two-dimensional (2D) materials and van der Waals (vdW) heterostructures. This process relies on the complementary pick-up and release temperatures ($T$) of the two polyvinyl chloride (PVC) films used here (PVC1: $T_{\text{pick up}} \approx 45\,°C$, $T_{\text{release}} \approx 90\,°C$; PVC2: $T_{\text{pick up}} \approx 90\,°C$, $T_{\text{release}} \approx 160\,°C$). (a) A vdW heterostructure, with 2D material sequence A/B/C, is assembled on the PVC1 working surface of Stamp 1 (e.g., see Figure 1). (b) This heterostructure is deposited by contacting a substrate at $T_{\text{release}}$ (90 °C for PVC1, shown here) and then slowly removing the stamp. (c) Alternatively, this heterostructure transfers from Stamp 1 (PVC1) to Stamp 2, with a PVC2 working surface, by contacting at 90 °C. Consequently, after the transfer, the sequence of flakes in the heterostructure is inverted relative to the stamp surface (C/B/A). (d) The flipped vdW heterostructure, now assembled on Stamp 2, is deposited by contacting a substrate at $\approx 160\,°C$. The glass slides seen in the photographs (b–d) are approximately 1 mm thick. The regions near the stamp and substrate surfaces are illustrated more clearly in the schematics but are not drawn to scale.

**Electrical Characterization of PVC-Assembled Heterostructures**

The physical characteristics of vdW heterostructures depend not only on the intrinsic properties of the constituent 2D materials but also on the details of their assembly. For example, extrinsic contaminants can affect carrier mobilities and charge doping levels in the active materials (e.g., graphene in the devices shown in Figure 3), causing fabricated devices to deviate from their idealized behavior. Typically, cleaner structures with more precisely controlled positions, orientations, and surroundings result in "higher quality" device characteristics. Many subtle, difficult-to-detect phenomena (e.g., quantum Hall effect in graphene) are easily masked by sample heterogeneity or contamination.[30,31]

We use low-temperature (≈ 40 K) electrical transport measurements to benchmark the quality of vdW heterostructure devices fabricated using the PVC-assisted assembly processes. Representative optical images of two devices are shown in Figure 3a,d. Device 1 (Figure 3a) is an hBN-encapsulated monolayer-graphene heterostructure (i.e., hBN/graphene/hBN), assembled and deposited using the process shown in Figure 1 (pick up and stack) and Figure 2a,b (deposit). Device 2 (Figure 3d) is a bilayer-graphene/hBN heterostructure that was stacked, flipped, and deposited (graphene side up) onto a prefabricated gold back gate electrode following the procedures shown in Figure 1 (pick up and stack) and Figure 2a,c,d (flip and deposit). Additional fabrication details are provided in Methods.

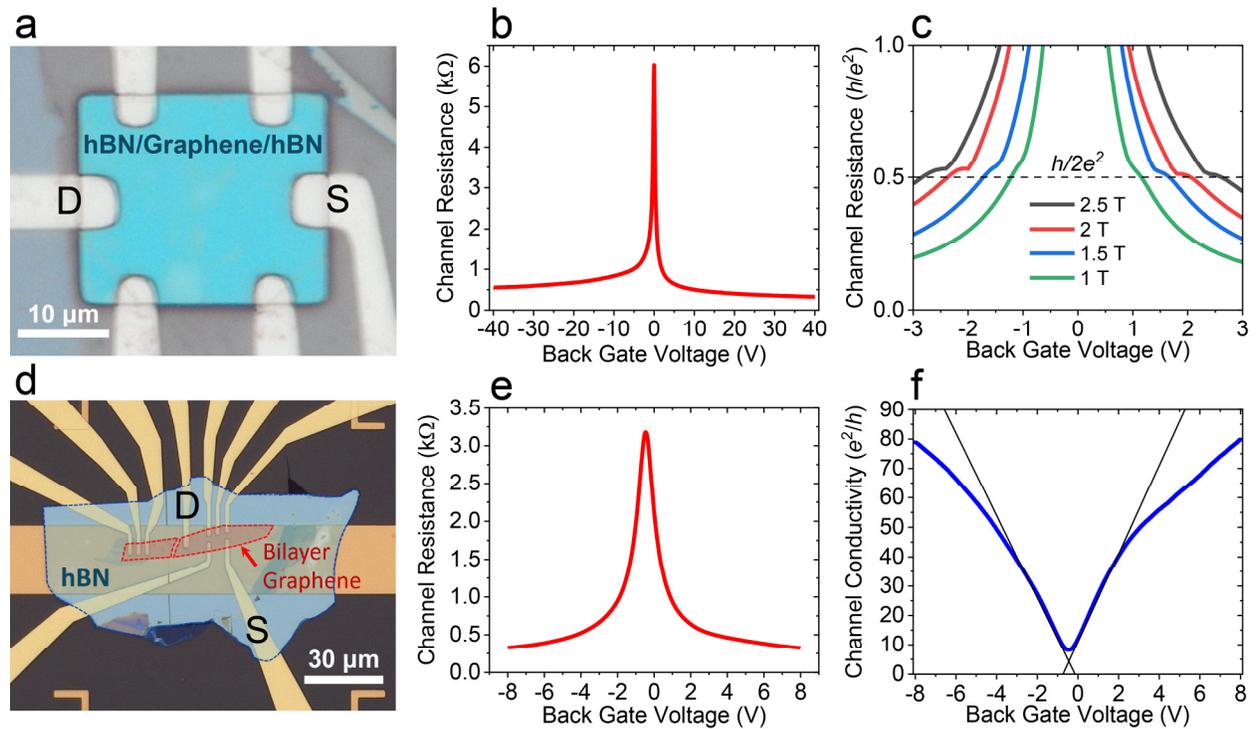

**Figure 3:** Characterization of hexagonal boron nitride (hBN) and graphene van der Waals (vdW) heterostructures assembled using polyvinyl chloride (PVC) stamps. (a) Optical image of an encapsulated hBN/monolayer-graphene/hBN device ("Device 1" in text), assembled using a PVC1 stamp, with metallic edge contacts and Si back gate. (b, c) Gate-modulated, two-terminal channel resistance of the device shown in (a), (b) at zero applied field, and (c) with an out-of-plane magnetic field. The quantum Hall plateaus at $h/(2e^2)$ (dashed horizontal line), starting at ≈ 1 T, are a clear indicator of high device quality. (d) Optical image of an exposed bilayer-graphene/hBN heterostructure ("Device 2" in text), stacked using a PVC1 stamp, then transferred to PVC2 stamp, flipped, and deposited onto prefabricated gold back gate electrodes. (e, f) Gate-modulated, two-terminal (e) resistance and (f) conductivity of the device shown in (c), illustrating the high sample quality with a virtually unshifted Dirac point. The slopes in the linear regions of the conductivity curve, here fit by black lines, estimate carrier mobilities. All electrical measurements were performed at ≈ 40 K inside a vacuum probe station using the labeled source (S) and drain (D) electrodes.

The electrical transfer characteristics of Device 1 (hBN/graphene/hBN) are shown, without magnetic field, in Figure 3b, and with magnetic field, in Figure 3c. The two-terminal resistance versus back gate potential ($V_{BG}$) of the encapsulated graphene device exhibits a sharp, virtually unshifted Dirac peak at $V_{BG} \approx 0$ V, indicating that there is no significant residual doping of the heterostructure. The carrier mobilities are estimated to be $\approx 5.5$ m$^2$ V$^{-1}$ s$^{-1}$ and $\approx 4.5$ m$^2$ V$^{-1}$ s$^{-1}$ for electrons and holes, respectively. Additionally, a quantum Hall plateau at $h/(2e^2)$ appears in the gate-modulated sample resistance when applying a relatively small (1 T) out-of-plane magnetic field, as seen in Figure 3c. The combination of $\approx 0$ V Dirac peak, high carrier mobilities, and the transition into quantum Hall transport at low magnetic field and relatively high temperature ($\approx 40$ K) all support that this device is one of the higher quality graphene samples to date.[32,33] Similarly, the gate-modulated two-terminal resistance of Device 2 (bilayer-graphene/hBN), shown in Figure 3e, exhibits a Dirac peak near zero back gate potential ($V_{BG} \approx 0.47$ V). Notably, however, Device 2 is not completely encapsulated and exposes the graphene bilayer at the top of the vdW heterostructure. This exposed surface leaves the graphene vulnerable to contamination effects from any residue coming from the stacking process and subsequent contact fabrication steps. Nevertheless, estimates of the carrier mobilities, derived from linear fits of the conductance curve, shown in Figure 3f, are $\approx 1.10$ m$^2$ V$^{-1}$ s$^{-1}$ (electron) and $\approx 1.05$ m$^2$ V$^{-1}$ s$^{-1}$ (hole). We note that these two-point measurements of the channel resistance, which include the metal/graphene contact resistances, systematically underestimate the carrier mobilities of both devices. Typically, contamination effects manifest as shifts in the Dirac point, higher-field transitions into quantum Hall transport, and lower carrier mobilities, due to proximity doping and increased carrier scattering. Here, both devices show excellent electrical characteristics that are indicative of insignificant residual contamination following the cleaning steps discussed below and in Methods.

**Residue Characterization & Sample Cleaning**

Stamp-assisted assembly techniques transfer contaminants from the polymer working surface to the contacted regions of 2D materials and the surrounding substrate.[34] The resulting heterogeneous layer of residue can interfere with subsequent measurements and device integration. In some circumstances, unwanted material can also become trapped beneath or between stacked flakes to form raised "bubbles" (or "blisters") of encapsulated contaminants.[35–37] In the case of PVC stamps, contamination is evident via optical microscopy and resolved more clearly in atomic force microscope (AFM) topographs (Figures 4a, 5, and S7). Similar contamination concerns exist for all polymer-assisted assembly techniques, as well as other processes used in device fabrication (e.g., lithographic patterning), with a range of remediation strategies reported.[38–40] Here, we characterize the contaminants found on surfaces contacted by PVC stamps and demonstrate effective cleaning strategies that enable high-performance device fabrication. Despite sample contamination being a significant concern for the fabrication of high-performance vdW devices when using any polymer-assisted assembly technique, careful examinations of the contaminants are rarely reported. We leverage high-resolution and compositionally sensitive methods to study the accidental transfer and removal of contaminants in detail. Such a detailed study, not previously reported for PVC or (to the best of our knowledge) for any common polymers used for heterostructure assembly, provides important insights for fabricating and cleaning high-quality 2D heterostructure devices.

The photothermal induced resonance (PTIR) technique, also known as AFM-IR, uses the probe tip of an AFM to transduce the photothermal expansion and contraction of a sample excited by a pulsed, tunable infrared (IR) laser.[41] Since the probe only transduces the sample excitations directly beneath the tip, this technique enables measurement of IR absorption spectra and maps with nanoscale spatial resolutions (≈ 10 nm),[42] far below the diffraction limit (several micrometers in the mid-IR). These spectra are comparable to far-field databases, which facilitate material identification[36,43] and semi-quantitative analysis[44] at the nanoscale in a wide range of applications[43,45] See Methods and recent reviews[41,46,47] for additional details.

Simultaneously acquired topography and PTIR absorption maps of an hBN flake deposited on a pre-patterned gold electrode using a PVC1 stamp are shown in Figure 4a and 4b, respectively. These measurements reveal the presence of numerous irregularly shaped, protruding features (contaminants) with increased absorption at 1735 cm$^{-1}$ relative to their surroundings. By holding the tip at fixed positions atop some of the larger (≈ 15 nm thick) contaminant features on the gold surface (points "3" and "4" in Figure 4), while sweeping the laser wavelength, we measure PTIR spectra (Figure 4c) that resemble Fourier transform IR absorption spectra of bulk PVC1 films (Figure S5). The two largest peaks in the PTIR spectra are observed at ≈ 1257 cm$^{-1}$ and ≈ 1735 cm$^{-1}$, which we attribute to H–C–Cl out-of-plane angular deformation and C=O stretching modes, respectively. Intuitively, we expect the H–C–Cl moiety from the PVC polymer chains, [CH$_2$–(H–C–Cl)–]$_n$, which comprise the stamp working surface, but this polymer lacks a C=O group. Therefore, the distinctive C=O stretch mode likely originates from one or more additives (e.g., phthalate or adipate plasticizers)[48,49] present in the PVC1 film. Since both modes are observed in the spectra, we conclude that both PVC and additives are transferred to the sample surface from the stamp. Accordingly, cleaning strategies, which have not been rigorously addressed in earlier works,[23–25] need to be developed to remove both these contaminants effectively. By contrast, on the transferred hBN flake, the topographic contaminant features are smaller (≲ 5 nm) and cover a lower fraction of the surface, in agreement with the weaker PTIR signal in this region. Spectra on the hBN flake (points "1" and "2") show a peak at ≈ 1368 cm$^{-1}$, which we attribute to the transverse optical phonon mode in hBN,[50,51] as well as barely detectable spectral features from the contaminants. These differences suggest lower abundances of the contaminants at these locations, in agreement with the topography and absorption maps. We note that quantitative comparisons of the spectra measured in the gold and hBN regions are not straightforward. In addition to the local contaminant abundance (proportional to the contaminant thickness),[44] the PTIR signal intensity depends on the plasmonic tip near-field[52] and the expansion of the underlying substrate.[53] The near-field enhancement is expected to be stronger on gold regions due to the smaller gap (≈ 15 nm) between the gold-coated tip and the gold substrate compared with the larger tip-substrate separation (≈ 110 nm) on the hBN-covered regions. Conversely, the thermal expansion coefficient of the hBN layer could boost the signal measured on the flake compared to the gold region.[53]

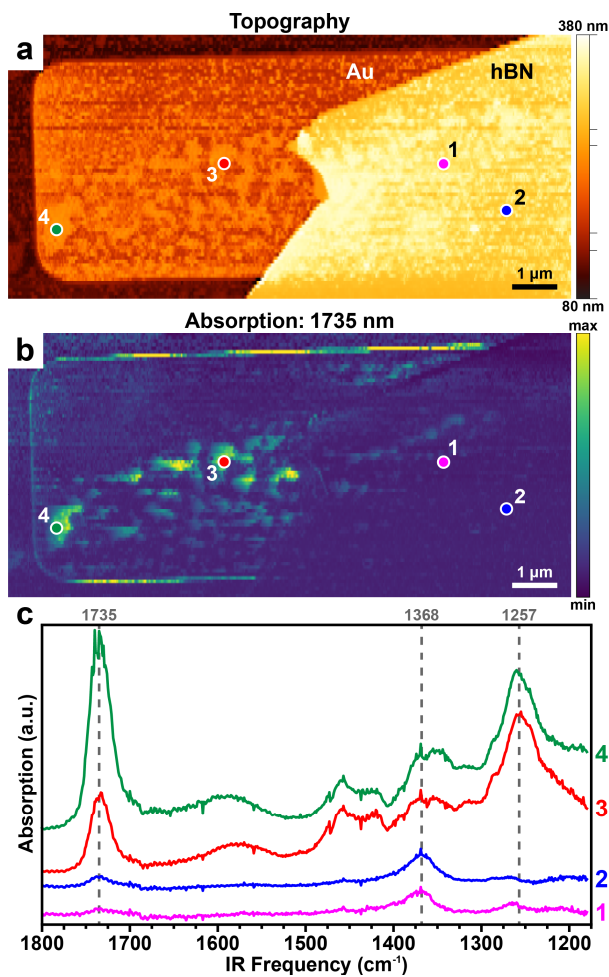

**Figure 4:** Characterization of polyvinyl chloride (PVC) residue around a hexagonal boron nitride (hBN) flake deposited on a gold electrode using an unconditioned PVC1 stamp. (a) Topographic map of the hBN flake (≈ 110 nm thick) and surrounding region directly contacted by the stamp reveals numerous small (typically < 15 nm in height), irregularly shaped contaminants. (b) Simultaneously acquired photothermal induced resonance (PTIR) absorption map of the same region shows enhanced infrared absorption at 1735 cm$^{-1}$, especially evident on some of the larger contaminants. (c) Peaks evident in PTIR absorption point spectra, measured at the locations indicated in (a) and (b), can be attributed to hBN (≈ 1368 cm$^{-1}$), PVC (≈ 1257 cm$^{-1}$, H–C–Cl out-of-plane angular deformation), and C=O (≈ 1735 cm$^{-1}$), which is not present in PVC but is a component in many common additives. Note that a nonlinear color scale is used in (a) to enhance the visibility of the contaminant topography; bands of approximately equal height change are indicated by tick marks along the right side with saturated regions below 80 nm and above 380 nm. Spectra in (c) are plotted using a common scale and vertically offset for clarity.

    The amount of residue transferred from a PVC stamp to another surface depends on the contact temperature. As illustrated in Figure S6, higher temperatures promote greater residue transfer from the stamp. To minimize this residue transfer, stamp/substrate temperature should not exceed the minimum temperature needed to deposit vdW assemblies reliably. For the commercial PVC films used here, we empirically determine these temperatures to be around 80 °C to 100 °C (PVC1) and around 150 °C to 160 °C (PVC2). Previously, Onodera and coworkers[23,24,54] observed

that pick-up temperature range corresponds with a glass transition in the polymer, while the temperatures used to release vdW structures correspond to the viscoelastic flow regime of their PVC/PDMS stamps.[23] These insights help to explain the temperature-dependent adhesion of PVC films and suggest strategies to engineer custom films (e.g., by varying film composition) with pick-up and release temperatures designed for specific applications.

Advantageously, the amount of residue transferred from a PVC stamp decreases with each additional surface contact, as illustrated in Figures 5 and S7. The greatest amount of residue transfers the first time the PVC surface contacts another substrate (e.g., to pick up a 2D flake). New, unused PVC1 stamps deposit numerous irregularly shaped contaminants during their first contact to a $SiO_2$ surface at 90 °C (Figure 5a,c). These contaminants are typically between about 100 nm and 200 nm in lateral extent and about 10 nm to 20 nm high. Subsequently, contacting the same PVC1 working surface to other bare regions of the substrate, for a second and third time, at 90 °C, transfers significantly less residue compared to the features found in the first-contact region. For example, contaminants appearing in the third-contact region are about ≈ 30 nm wide and ≲ 5 nm high. This trend is also observed for stamps made with fresh PVC2 working surfaces (Figure S7a,c), with less residue transferred from the stamp with each additional contact. Conditioning contacts, such as these, before 2D material transfer and assembly, at temperatures that match or exceed typical flake-release temperatures, is therefore an effective strategy to reduce the amount of PVC residue transferred from the stamp during subsequent contacts. Importantly, we note the PVC films employed here can contact surfaces (e.g., for conditioning, pick-up, and deposition) many times without significant physical degradation. In fact, as evidenced by the trends in the amounts of transferred residue, PVC stamps become *cleaner* with increased use (i.e., surface contact), though with diminishing returns. In stark contrast to these PVC films, other polymers (e.g., PPC), require repeated annealing to relieve stress in the working surface after each contact step, typically only lasting a few cycles before becoming unusable. These limitations disrupt the assembly process and constrain the sizes and complexities of the structures that can be produced using other polymers, making PVC films a favorable alternative for 2D material assembly.

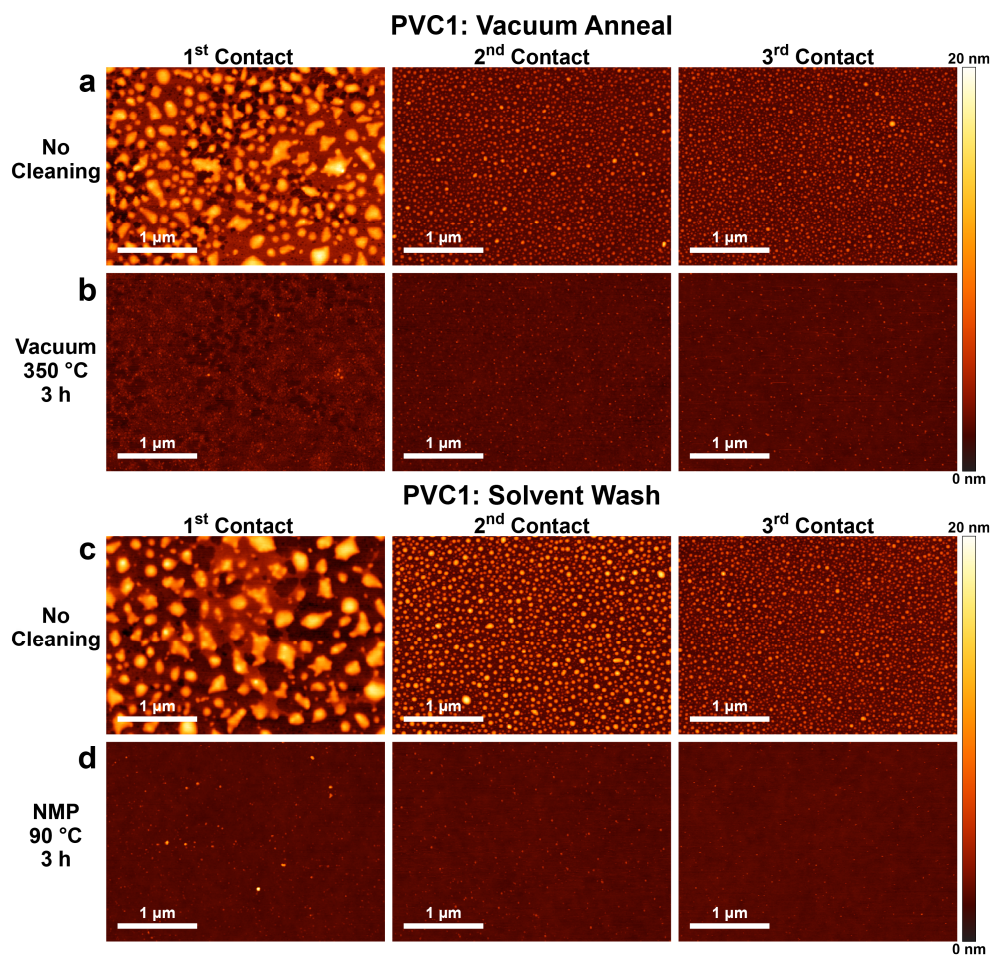

**Figure 5:** Atomic force topographs of $SiO_2$/Si substrates contacted by polymer stamps with bare PVC1 working surfaces (i.e., no two-dimensional materials). Images in the same row compare different regions of the substrate contacted sequentially by the same stamp, indicated with ordinal numbers: 1$^{st}$, 2$^{nd}$, 3$^{rd}$, at 90 °C. (a, c) After removing the stamps, the contacted regions of the substrates are nearly covered with irregularly shaped contaminants transferred from the PVC1 surfaces. Contaminants are most prominent in the first-contact regions, containing the largest-area, most-protruding surface features. Progressively less contamination is observed in the second- and third-contact regions. (b, d) These contaminants are substantially removed after cleaning with either the vacuum annealing or solvent washing protocols described in the text. (b) The same regions shown in (a) after annealing in vacuum at ≈ 350 °C for about 3 h. (d) The same regions shown in (c) after washing in N-methyl-2-pyrrolidone (NMP) at ≈ 90 °C for about 3 h. To enable easier side-by-side comparisons, the height (color) scale in all images is set to a common range (20 nm), with the plane of the substrate approximately equalized in each panel. Regions outside of this height range are shown saturated at the limits displayed. See also Figures S7 and S8.

Despite significantly reducing the amount of transferred residue, conditioning PVC stamps before 2D material assembly does not entirely prevent contamination, thus motivating additional strategies to obtain residue-free vdW heterostructures. Here, we demonstrate two post-fabrication cleaning strategies (vacuum annealing and solvent washing) to remove PVC residue. These strategies have different merits and use cases that depend on the specific characteristics and requirements of the fabricated structures. High-temperature vacuum annealing (or, alternatively, annealing in a suitable gas, e.g., Ar or $H_2$) breaks down contaminants into more volatile species that are removed from the surface without directly contacting the delicate assemblies.[55,56] However, some 2D materials (e.g., $Bi_2Sr_2CaCu_2O_{8+x}$[57] and $MoO_3$[42]) are unstable and degrade at typical annealing temperatures (e.g., $\geq 350\ °C$). Further, vdW heterostructures with precisely controlled interlayer twists (e.g., "magic angle" graphene or other moiré superlattices[26]) can relax into more stable twist configurations at elevated temperatures, which alters the carefully engineered properties of the assembly. In contrast to vacuum annealing, solvent-based cleaning techniques can remove contaminants at lower temperatures (e.g., between about 20 °C and 100 °C for many common solvents), but are ineffective towards buried contaminants and can inadvertently delaminate (i.e., "wash away") vdW heterostructures from the surface. Nevertheless, depending on the material characteristics, one or both of these strategies can be used to clean a wide range of vdW assemblies of interest to researchers.

We demonstrate the effectiveness of vacuum annealing at removing PVC residue from $SiO_2$ surfaces. Figure 5 directly compares the same regions of an otherwise bare $SiO_2$/Si substrate contacted by a PVC1 stamp before (contaminated, Figure 5a) and after (cleaned, Figure 5b) annealing in a vacuum at 350 °C for 3 h. After annealing, the previously contaminated regions appear substantially cleaner, with the topographic features observed beforehand significantly reduced in size. As discussed above, the amount of contamination depends on the number of times the PVC stamp has previously made contact with the substrate and the contact temperature. Accordingly, of the three regions compared, the first-contact region (at 110 °C) exhibits the largest contaminant features, which have irregular but identifiable shapes. After annealing, features with similar shapes and positions to the initial contaminants can still be seen in the surface topography, though protruding significantly less ($\lesssim 1$ nm) above the surrounding substrate compared to the features observed before annealing ($\approx 10$ nm). In the cases of the initially less-contaminated second- and third-contact regions, the contaminants that remain after annealing are much less apparent. We note that, after annealing, numerous small "speckle" features (less than a few nanometers in height and tip-width limited in diameter) are observed across all three regions. The origin of these features after annealing is unclear as they are also observed in regions never contacted by PVC stamps (Figure S8). Nevertheless, in large part, vacuum annealing effectively removes most PVC1 contaminants from the second- and third-contact regions of the surface, which also bolsters the case of using one or more conditioning contacts prior to assembling 2D materials with PVC stamps. Although less effective, lower annealing temperatures (e.g., 250 °C to 350 °C) or shorter durations could also be used to remove residual polymer contaminants from less thermally robust structures. However, cleaner surfaces are typically obtained by annealing at higher temperatures ($\geq 350\ °C$)[55,58] for longer times, when viable.

Alternatively, washing vdW heterostructures in solvents removes contaminants without subjecting them to high temperatures or producing non-volatile decomposition products that remain after annealing. Several solvents, including tetrahydrofuran, dimethylformamide, and various ketones can dissolve PVC.[59,60] Nevertheless, in practice, these solvents are not equally effective at removing PVC-derived residue and can leave minute traces on the surface after washing. Longer solvent immersion times typically result in more complete contaminant removal. Additionally, although washing with room-temperature solvents is the simpler approach, mild heating (below their respective boiling points) increases contaminant removal rates and overall cleaning effectiveness.[61] For example, following the stack-and-flip assembly and deposition used to fabricate Device 2 (Figure 3d), the graphene/hBN heterostructure was immersed in cyclohexanone at ≈ 70 °C for ≈ 2 h, and then left in the solvent overnight (≈ 10 h at ≈ 20 °C), to remove the PVC-derived residue.

Among the solvents tested in this work (see Methods), washing with N-methyl-2-pyrrolidone (NMP) is found to be the most effective at removing PVC-derived residue. Figure 5c,d compares the same regions of a $SiO_2$/Si substrate contaminated by contact with a PVC1 stamp before and after washing in NMP at ≈ 90 °C for 3 h. After washing, the previously contaminated regions appear significantly cleaner and possess essentially no residue with only a few small, particle-like features. We note that these same features remain after a subsequent vacuum annealing process (450 °C for 3 h; not shown), suggesting that they are not typical organic residue. Instead, these thermally resilient, insoluble contaminants may be other material(s) that are adventitiously present on the film and are readily removed by conditioning contacts; see Methods.

Similar comparisons are provided in Figure S7 for substrates contacted by PVC2 stamps, illustrating analogous trends in vacuum annealing and solvent washing effectiveness. In general, residue derived from PVC2 is found to be more difficult to clean than residue from PVC1, which we hypothesize is related to differences in the film compositions (see Methods and the Supplementary Information).

**Heterogeneous Assembly of Photonic Devices**

To highlight the versatility of PVC-assisted assembly, we extend this emerging technique beyond 2D materials. As an additional demonstration, we use PVC1 stamps to pick up and deposit nanolithographically patterned slabs of aluminum gallium arsenide ($Al_xGa_{1-x}As$; here $x \approx 0.4$). This bulk, three-dimensionally (3D) structured semiconducting material exhibits non-linear optical properties with a composition-tunable bandgap and refractive index, which is useful for many integrated photonic applications.[62–64] Typically, $Al_xGa_{1-x}As$ is pattered into optoelectronic devices using traditional CMOS-compatible nanofabrication techniques. However, heterogeneous integration[65,66] of photonic components made from different material technologies (e.g., combining III-V materials, such as GaAs, with Si or $Si_3N_4$) sometimes requires removing the components from their original substrates.[67,68] Here, we pick up patterned $Al_{0.4}Ga_{0.6}As$ structures (analogous to many other types of CMOS-fabricated devices) and controllably deposit them onto a prefabricated $Si_3N_4$ optical resonator cavity using a PVC stamp. This capability enables the creation of heterogeneously integrated photonics devices and, more broadly, hybrid 2D/3D heterostructures that would be challenging to produce by other means.

Following the procedure used for 2D material transfer, (e.g., Figure 1), we pick up $Al_{0.4}Ga_{0.6}As$ ("bullseyes") structures from a nanolithographically patterned substrate. Importantly, to facilitate removing these structures made from epitaxially grown bulk materials that strongly adhere to the underlying substrate, a cavity is etched beneath each structure (see Methods).[64] This cavity partially releases the structures from the substrate, leaving them suspended and tethered by 4 spokes to the semi-continuous $Al_{0.4}Ga_{0.6}As$ film on top (see Figures 6a and S9a). During the pick-up process, a PVC1 stamp contacts and compresses the patterned surface, causing the suspended $Al_{0.4}Ga_{0.6}As$ nanostructures to buckle and then break at their tethers. Lifting the stamp picks up the now-fully released nanostructures, which adhere to the PVC1 surface at ≈ 45 °C. Analogous to the 2D material transfer demonstrated above (e.g., Figure 2a,b), these nanostructures are deposited by contacting them to the target surface at ≈ 90 °C and then slowly removing the stamp. Figure 6b shows the transferred $Al_{0.4}Ga_{0.6}As$ nanostructures deposited near the edge of a $Si_3N_4$ optical resonator. These demonstrations illustrate the wide-ranging versatility of PVC-assisted materials assembly, which includes both 2D as well as bulk, 3D structured nanomaterials (see Table S1 for a summarized list).

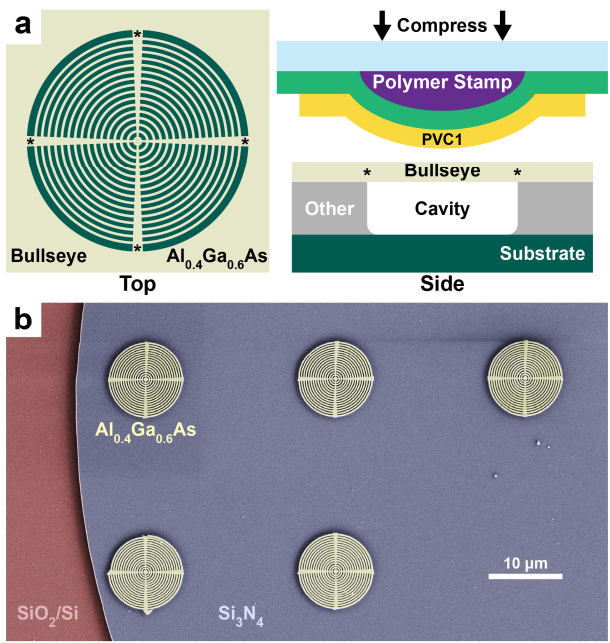

**Figure 6:** Transfer of bulk, three-dimensionally structured materials using polyvinyl chloride (PVC) stamps. (a) Schematics showing top-down (left) and side (right) views of an $Al_{0.4}Ga_{0.6}As$ "bullseye" nanostructure ($\approx$ 135 nm thick), suspended over an air-filled cavity and weakly attached by tethers to the surrounding $Al_{0.4}Ga_{0.6}As$ film out of which it is patterned. Contacting and compressing the suspended structure with a polymer stamp from above strains the tethers, which break near the locations indicated by *. Afterwards, the stamp picks up (at $\approx$ 45 °C) the fully released structure and transfers it to another substrate. (b) False-color scanning electron micrograph of 5 simultaneously transferred $Al_{0.4}Ga_{0.6}As$ nanostructures, deposited (at $\approx$ 90 °C) atop a $Si_3N_4$ optical resonator using a PVC1 stamp. The transferred nanostructures are shown after vacuum annealing at $\approx$ 400 °C for 10 h. See also Figure S9.

## Conclusion

In this work, we assemble 2D materials and vdW heterostructures using polymer stamps made from commercially available PVC films and confirm their high-performance characteristics. Commercial films, which are mass-produced with consistent characteristics and sold in large, inexpensive rolls, greatly simplify polymer stamp construction by eliminating the preparation of the custom polymer films typically used for 2D material assembly. Accordingly, researchers can treat these easily replaceable polymer films akin to other materials used to construct stamps (e.g., glass slides and tape). Compared with other polymers (e.g., PPC), PVC films are significantly more durable. This durability enables reliable pick up and deposition of a wide variety of 2D materials over many pick-up and release cycles, thereby facilitating fabrication of complex devices. The complementary pick-up and deposition temperatures of the two PVC films employed here enable P2P, stack-and-flip 2D materials assembly that is otherwise prohibitively challenging to achieve through other means. We thoroughly characterize polymer contaminants transferred from the stamps and develop effective cleaning strategies to produce atomically clean interfaces. Electrical transport measurements of graphene and hBN vdW heterostructures assembled using PVC stamps verify the high-quality characteristics of the materials, including virtually unshifted Dirac peaks (≈ 0 V), high carrier mobilities, and the emergence of quantum Hall conduction at low magnetic field. Additionally, for the first time, we use PVC1 stamps to pick up and to deposit bulk $Al_{0.4}Ga_{0.6}As$ nanostructures onto $Si_3N_4$ optical resonators to produce heterogeneously integrated photonic devices. This capability enables the creation of hybrid structures and devices that combine different materials technologies assembled using the same, highly versatile technique. The significant advances demonstrated here enable broader adoption of PVC-assisted assembly and facilitate rapid fabrication of complex vdW heterostructures and nanomaterials assemblies for a wide range of applications.

## Methods

### Materials

Commercial PVC films were obtained from Sign Outlet Store (Lisle, IL, USA),[69] "Series 3200 Optically Clear overlaminate," (≈ 50 µm, PVC1), and from McMaster-Carr,[70] "Clear Chemical-Resistant PVC Film," (8562K11, ≈ 100 µm thick, PVC2).[a] These two films were produced by Arlon Graphics (Placentia, CA, USA) and Warp Brothers/Flex-O-Glass, Inc. (Chicago, IL, USA), respectively.[a] Both PVC films arrived packaged as long (≈ 90 cm) cylindrical rolls, which were cut into shorter rolls (≈ 1.5 cm wide) before use. We note that the safety data sheet (SDS) for PVC2 indicates the films contain between 19 % and 32 % di(2-ethylhexyl) phthalate and diisononyl phthalate plasticizers. Other additives indicated by the PVC2 SDS include 1 % to 3 % metallic (Zn, Ba) stearates, used as thermal stabilizers and siloxane and styrene related polymers that comprise < 10 % of the total.

### Stamp Construction

Stamp working surfaces consist of thin films of either PVC1 or PVC2 supported by a dome-shaped scaffold made from PDMS and transparent tape (Scotch Transparent Tape).[a] The scaffolds are produced by cutting slabs of PDMS (approximately 1 mm × 1 mm × 0.5 mm) from Gel-Pak4 Gel-Boxes and positioning them near the ends of transparent cantilevers (glass slides).[a] The PDMS slabs are taped to the glass surface and squeezed into rounded domes by pressing and sliding obliquely angled razor blades across the tape surface, particularly near the corners of the PDMS slab; see schematic (Figure S1b) and photographs (Figure S2) of stamp construction in the Supplementary Information. Importantly, an adhesive coating on the back side of PVC1 films helps to attach it securely when pressed onto the scaffold. By contrast, PVC2 films do not possess this adhesive backing. Instead, for PVC2 stamps, a layer (not shown in Figure S1) of clear, heat-resistant, double-sided polyethylene terephthalate tape replaces the single-sided tape used to form the dome-shaped scaffold. Without a strong attachment to the scaffold, repeated contacts between the stamp and sample substrates can cause deformation and delamination of the PVC2 working surfaces, which negatively impacts control over the location and size of the contact region.

Since all the materials used in the stamp construction are optically transparent, they do not prohibitively obscure the view of a microscope focused on a sample surface during 2D flake manipulation. Consequently, when combined with an adjustable stage, the position and orientation can be readily controlled in vdW heterostructure assemblies.

**Stamp Conditioning**

Conditioning PVC stamps substantially reduces contamination of the deposited structures. Before 2D material assembly, PVC stamps are conditioned by directly contacting a clean $SiO_2$/Si substrate at, or above, the release temperatures for the respective PVC films (at 110 °C for PVC1 stamps or at 160 °C for PVC2 stamps). During conditioning, the stamps are kept in contact with the substrate for ≈ 1 min and then separated. Only regions of the stamp working surface that make direct contact with the substrate are conditioned in this process. Typically, this conditioning process is repeated 2 to 3 times on separate, clean regions of a $SiO_2$/Si substrate. We note, however, that stamp conditioning was *not* performed before assembling Device 1 and Device 2, nor before transferring the hBN flake shown in Figure 4.

**Device Fabrication**

Device 1 was fabricated on $SiO_2$ (≈ 290 nm thick) thermally grown atop a heavily doped $p^+$Si substrate, which acts as a back gate. Exfoliated hBN (≈ 100 nm thick) and monolayer graphene flakes were assembled using a PVC1 stamp, and the resulting hBN/graphene/hBN heterostructure was deposited onto pre-patterned alignment marks following procedures shown in Figures 1 and 2a,b. The heterostructure was then patterned into a conventional Hall bar geometry by using photolithography and subsequent reactive ion etching. A second photolithography step, followed by deposition of Cr/Pd/Au (5 nm/20 nm/95 nm) and a subsequent metal lift-off process, was used to create one-dimensional metal edge contacts[32] to the encapsulated graphene. By contrast, Device 2, a bilayer-graphene/hBN (65 nm) heterostructure, was deposited onto a pre-patterned gold electrode (back gate) following procedure shown in Figures 1 and 2a,c,d.

The deposited heterostructures were cleaned in heated (≈ 70 °C) cyclohexanone for ≈ 30 min, rinsed with isopropanol (IPA), and blown dry using nitrogen. After this solvent washing, Device 2 was annealed in a vacuum (pressure ≈ $10^{-4}$ Pa) at ≈ 350 °C (ramp rate of ≈ 0.2 °C/s) for about 3 h to remove any remaining residue. Additional bulk electrical contacts were fabricated following procedures that were developed previously.

Throughout this work, stamp contact temperatures are considered approximate, nominal values. In practice, uncertainties in stamp contact temperature depend on the configuration of the stacking stage assembly (see Figure S1), e.g., due to differences in the positions of the heater and temperature sensor, as well as the type and thickness of the contacted substrate.

**Scanning Probe Measurements**

All PTIR measurements in this work use a modified, commercially available instrument operated in resonance-enhanced contact mode, as described previously.[36,72] Briefly, a quantum cascade laser with a tunable pulse repetition rate (1 kHz to 2000 kHz) and wavelength (910 cm$^{-1}$ to 1905 cm$^{-1}$) illuminated the sample. The laser light (p-polarized) focused to a spot, ≈ 50 µm in diameter, centered around the tip of a gold-coated Si probe (nominal spring constant 0.07 N/m to 0.4 N/m and first resonance frequency in air of 13 kHz ± 4 kHz). A phase-locked loop maintained the pulse repetition rate at one of the contact-resonance modes to achieve a resonance signal enhancement proportional to the quality factor of the probe.

All other surface topography measurements (outside the context of PTIR) were acquired using a commercial AFM operated in tapping mode using an aluminum-coated probe with nominal spring constant between 20 N/m and 75 N/m and first resonance frequency in air of in the range of 200 kHz to 400 kHz.

**Solvent Washing**

Solvents tested in this work to remove PVC-derived residue include tetrahydrofuran, dimethylformamide, cyclohexanone, cyclopentanone, acetone, chloroform, toluene, hexane, IPA, and NMP. All solvents were acquired from commercial sources and used as received. Additionally, we found that Remover PG (a proprietary, NMP-based solvent designed to remove resists from Si and $SiO_2$ surfaces)[a] was also extremely effective at removing PVC-derived residue. However, Remover PG also demonstrated a tendency to cause delamination of deposited heterostructures from surfaces, especially when used continuously for long periods of time (e.g., overnight). We also note that the washing protocols described in this work do not necessarily represent the optimal strategy (e.g., solvent temperature, duration, etc.) in all use cases.

**Fabrication of Photonics Devices**

The $Si_3N_4$ photonics and $Al_{0.4}Ga_{0.6}As$ structures used in this work are fabricated using previously published methods.[64,73] We note that the released "bullseye" nanostructures are fabricated from an original film stack composed of $SiO_2$ (< 100 nm), GaAs (≈ 5 nm), $Al_{0.4}Ga_{0.6}As$ (≈ 140 nm), GaAs (≈ 5 nm), $Al_{0.75}Ga_{0.25}As$ (≈ 500 nm), GaAs (bulk substrate). After etching the $Al_{0.75}Ga_{0.25}As$ intermediate layer, thereby releasing the structures from the underlying GaAs substrate, the suspended film consists of a slab of $Al_{0.4}Ga_{0.6}As$ sandwiched between two thin layers of GaAs on the top and bottom surfaces. For simplicity, the composition of these structures is simply referred to as $Al_{0.4}Ga_{0.6}As$ in the text.

## Data availability statement

All data that support the findings of this study are included within the article (and any supplementary files).

## Notes

[a]The full description of the procedures used in this paper requires the identification of certain commercial products and their supplier. The inclusion of such information should in no way be construed as indicating that such products are endorsed by NIST or are recommended by NIST or that they are necessarily the best materials for the purposes described.

## Conflicts of interest

The authors declare no competing financial interests.

## Acknowledgments

The authors acknowledge resources provided by the NanoFab at the National Institute of Standards and Technology (NIST) and are grateful for assistance from Doug Ketchum, Edgar Perez, and Ramesh Kudalippalliyalil (Laboratory for Physical Sciences), Sadhvikas Addamane (Sandia National Labs), Marcelo Davanco (NIST), Junyeob Song and Ashish Chanana (Theiss Research and NIST), and Will Eshbaugh (West Virginia University and NIST).

*Supplementary Information for*

# Assembly of High-Performance van der Waals Devices Using Commercial Polyvinyl Chloride Films

Son T. Le,[1,2,*,#] Jeffrey J. Schwartz,[1,2,3,*,#] Tsegereda K. Esatu,[1,2] Sharadh Jois,[2]

Andrea Centrone,[3] Karen E. Grutter,[2] Aubrey T. Hanbicki,[2] Adam L. Friedman[2,*]

[1]Department of Electrical and Computer Engineering, University of Maryland, College Park, Maryland 20742, United States

[2]Laboratory for Physical Sciences, College Park, Maryland 20740, United States

[3]Physical Measurement Laboratory, National Institute of Standards and Technology, Gaithersburg, Maryland 20899, United States

[#]S.T.L. and J.J.S. contributed equally

**Contents:**

Figure S1: Schematic illustrations of the fabrication and use of a polymer stamp

Figure S2: Photographs showing the step-by-step construction of polymer stamps

Figure S3: Photographs of two-dimensional (2D) materials transfer and assembly using a polymer stamp

Figure S4: Photographs of polymer-to-polymer transfer and flipping of two-dimensional (2D) materials using polyvinyl chloride (PVC) stamps

Vibrational Spectroscopy of Polyvinyl Chloride

Figure S5: Vibrational spectroscopy of polyvinyl chloride films

Figure S6: Atomic force topographs of $SiO_2$/Si substrates contacted at different temperatures by PVC1 stamps

Figure S7: Atomic force topographs of $SiO_2$/Si substrates contacted by PVC2 stamps and cleaned

Figure S8: Atomic force topographs of $SiO_2$/Si substrates uncontacted by PVC stamps and cleaned

Figure S9: Controlled pick up and deposition of $Al_{0.4}Ga_{0.6}As$ nanostructures to create a heterogeneously integrated photonic device

Table S1: List of materials tested with PVC-assisted pickup and deposition

Supplementary References

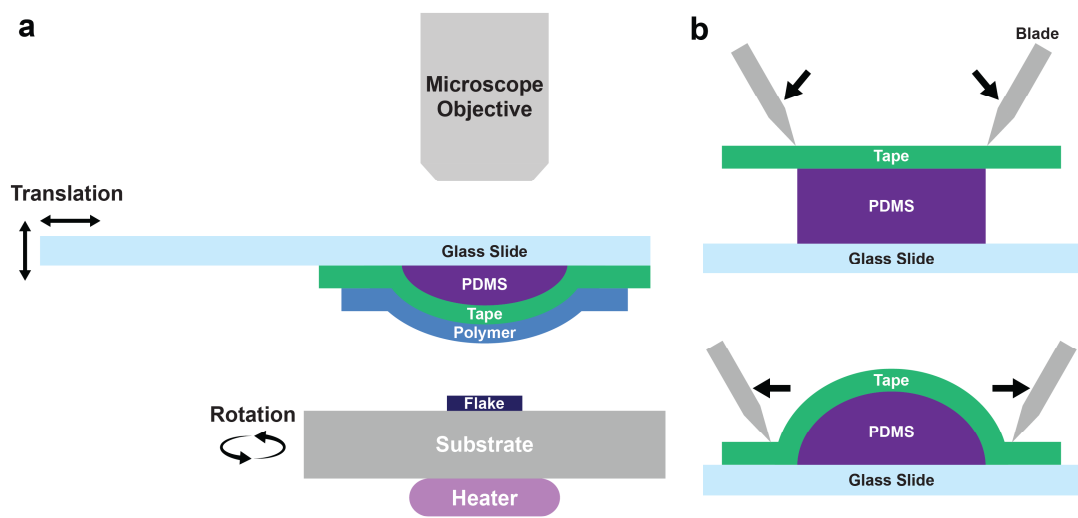

**Figure S1:** Schematic illustrations of the fabrication and use of a polymer stamp for the transfer and assembly of two-dimensional (2D) materials. (a) Stamps consist of a working polymer film supported by a dome-shaped scaffold, made from polydimethylsiloxane (PDMS) and transparent tape, positioned near the end of a glass cantilever. During use, the stamp is suspended beneath a microscope objective by a translation stage (not shown) that can move the cantilever both normal to and in the plane of the sample substrate. The substrate is positioned below the stamp on a rotation stage (not shown) with a built-in heater for temperature control. (b) The scaffold supporting the working polymer is constructed using transparent tape to seal a thin slab of PDMS to the glass surface. Using razor blades, the tape is pressed down at the edges of the slab and adheres to the glass surface (bottom schematic). Tension from tape, stretched between the top of the PDMS slab and the underlying glass, continues to exert stress on the PDMS, deforming and rounding its profile. By repeating this process at each edge, the initially rectangular PDMS slab takes on a dome-like shape. The stamp working surface, supported above the apex of this scaffold, is the first portion of the stamp to contact the sample substrate. Sharper profiles (i.e., smaller radii of curvature) can enable more precise manipulation of 2D flakes than flatter profiles (larger radii of curvature) but are also subject to more optical distortions when viewing the sample due to refraction through the stamp. Note that these schematics are not depicted to scale. See also Figure S2.

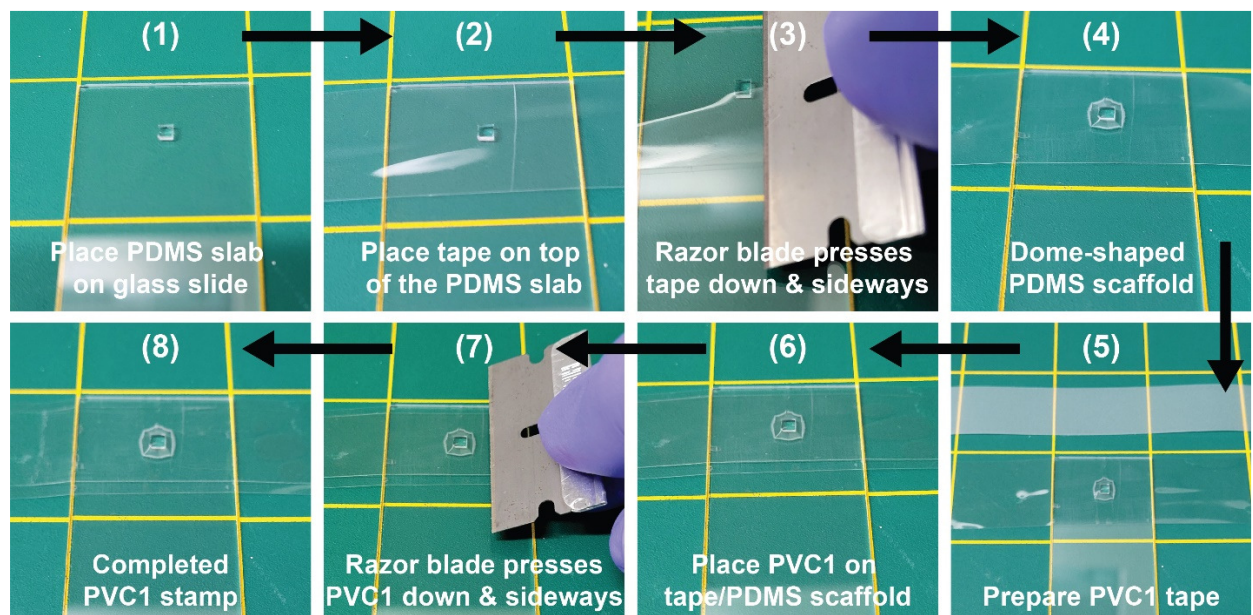

**Figure S2:** Photographs showing the step-by-step construction of polymer stamps with PVC1 working surfaces. (1) A slab of polydimethylsiloxane (PDMS), measuring approximately 1 mm × 1 mm × 0.5 mm, is positioned near the end of a glass slide. (2) Transparent tape is placed gently on top of the PDMS slab. (3) A razor blade presses the tape down, toward the glass and away from the PDMS, at the edges of the slab. (4) After pressing the tape to the glass along each side of the PDMS slab, the tape squeezes the slab into a dome shape (scaffold). See also Figure S1b. (5) A strip of PVC1 tape (with adhesive back coating) is cut to size and its protective backing film is removed. (6) The PVC1 tape is placed gently on the dome-shaped scaffold (analogously to step 2) and (7 and 8) pressed down using a razor blade, similar to steps 3 and 4. Afterward, excess material on the sides of the stramp can be trimmed away using razor blade. Constructing PVC2 stamps is similar to the processes described here but adds a layer of transparent, double-sided tape over the bottom transparent tape (after steps 1 – 4) to compensate for the lack of an adhesive coating on PVC2 films. For scale, parallel yellow lines on the underlying work surface (green background) are separated by 2.54 cm.

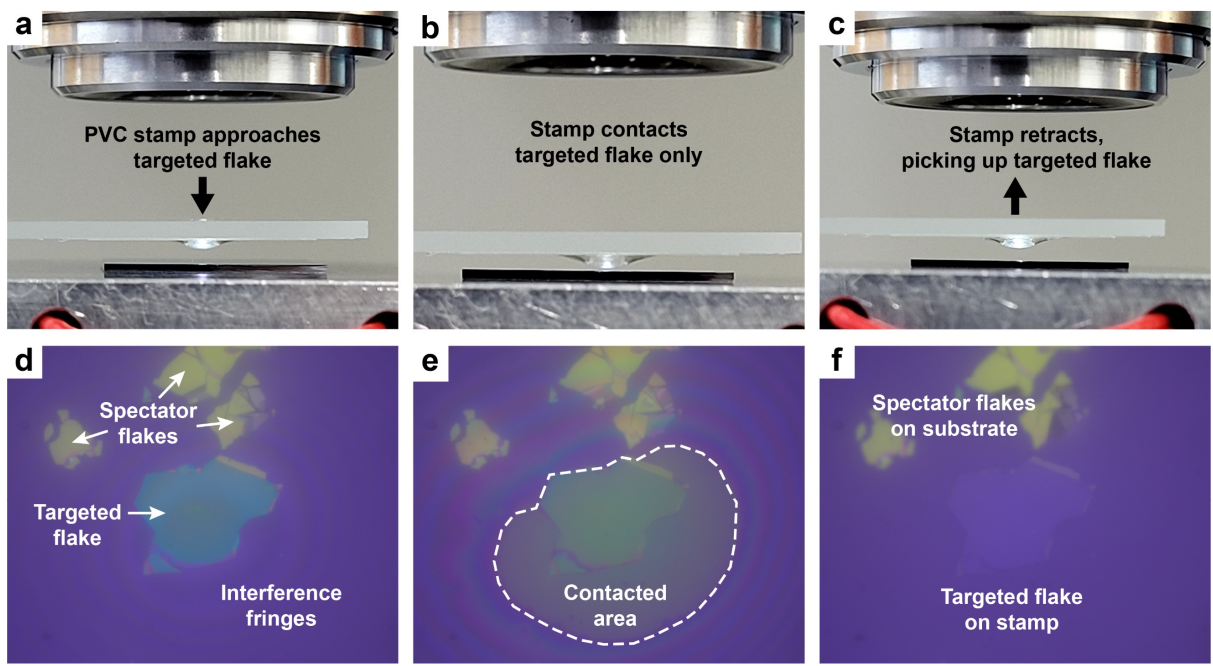

**Figure S3:** Photographs of two-dimensional (2D) material transfer and assembly using a polymer stamp. (a–c) Macroscopic side view of the 2D flake pick-up process. (a) The polyvinyl chloride (PVC) stamp (a) approaches and (b) contacts the targeted 2D flakes at the pick-up temperature ($T_{\text{PVC1 pick up}} \approx 45\ °C$), maintained by the heated stage. (c) After contact, the stamp is slowly removed from the substrate, in the process picking up the targeted flake. For scale, the glass slide used to construct the stamp is approximately 1 mm thick. (d–f) Microscopic top-down views of the 2D flake pick-up process. (d) Several exfoliated 2D flakes of different thicknesses and lateral sizes are visible on a $SiO_2$/Si substrate. The largest flake is targeted in the transfer process shown here. When the stamp is very near the substrate (a few micrometers or less), interference fringes (rings) become visible. The colors, intensities, and shapes of these fringes provide an indication of the stamp-substrate separation distance. Furthermore, the center locations and shapes of these rings also give insight into the location and shape of the apex of the dome-shaped stamp working surface that will contact the substrate. (e) Upon contact, the working surface deforms slightly as a contact wavefront (dashed lines) expands across the surface to cover the target flake. (f) Lifting the stamp transfers the flake from the substrate to the PVC working surface. Picking up the thin flake from the substrate typically results in a change in its apparent color. As the stamp moves farther from the substrate, flakes located on the stamp and those remaining on the substrate will no longer share the same focal position when viewed through a microscope objective. For scale, the targeted flake is approximately 70 µm in its overall length and width. See also, analogously, the schematic process shown in Figure 1.

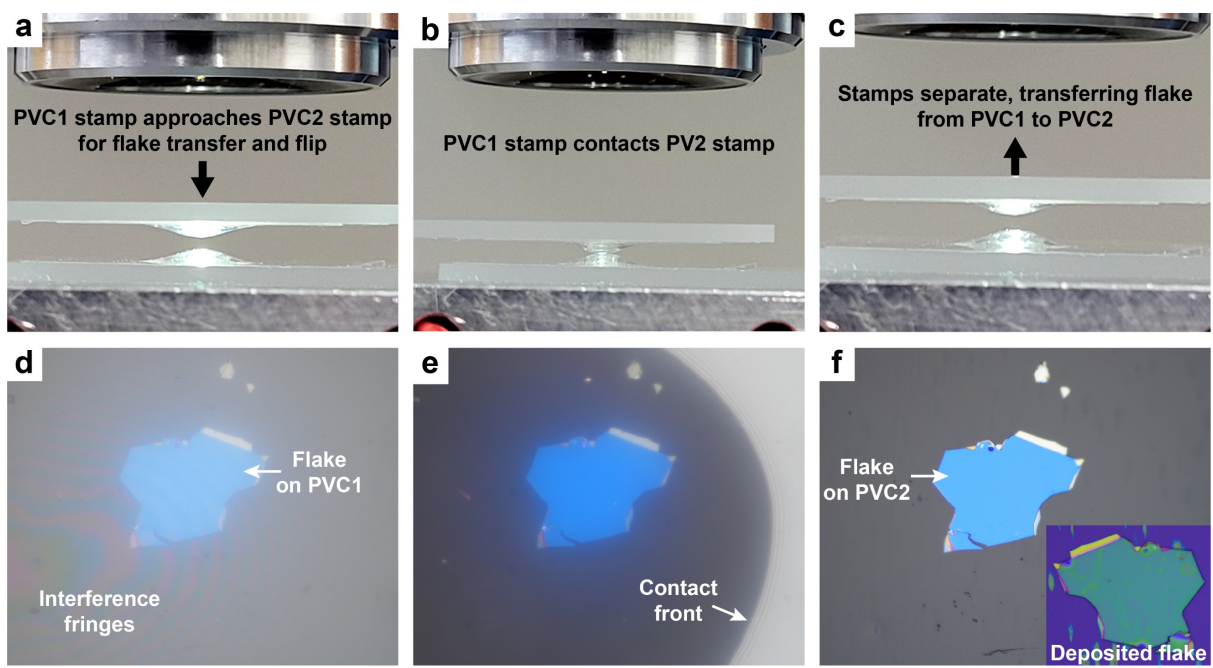

**Figure S4:** Photographs of polymer-to-polymer transfer and flipping of two-dimensional (2D) materials using polyvinyl chloride (PVC) stamps. (a–c) Macroscopic side view of the process that transfers a flake between two polymer stamps. On top, a PVC1 stamp carrying the 2D flake (a) approaches and (b) contacts the PVC2 stamp, on bottom. At ≈ 90 °C, corresponding to the approximate release temperature of PVC1 and also the pick-up temperature of PVC2 ($T_{\text{PVC1 release}} \approx T_{\text{PVC2 pick up}} \approx 90$ °C), the 2D flake is released from the PVC1 stamp while simultaneously adhering to (picked up by) the PVC2 stamp. (c) Separating the stamps leaves the flake, now flipped relative to the contacting PVC surface, on the PVC2 stamp and available for subsequent deposition onto another substrate. For scale, the glass slide used to construct the stamp is ≈ 1 mm thick. (d–f) Microscopic top-down views of the transfer process. (d) When the two stamps are in close proximity (a few micrometers or less), interference fringes (rings) become visible. The colors, intensities, and shapes of these fringes provide an indication of the stamp-stamp separation distance. Furthermore, the center locations and shapes of these rings also give insight into the locations and shapes of the apexes of the dome-shaped stamp working surfaces that will make contact. (e) Upon contact, the working surfaces deform slightly as the contact wavefront expands. (f) Separating the stamps transfers the flake from the PVC1 surface to the PVC2 surface. Note that the flake in (f) is supported from below by the PVC2 stamp and is viewed from above, so it appears the same as in the previous images that also view the flake from above by looking through the transparent backside of the PVC1 stamp. After the transfer, however, the flake is effectively flipped upside-down relative to the initial PVC1 stamp since the flake is supported on its opposite face by the PVC2 stamp. (Inset) After the transfer, the flake can be deposited onto another substrate from the PVC2 stamp, viewed here in its final, flipped orientation. For scale, the flake is approximately 70 µm in its overall length and width. See also, analogously, the schematic process shown in Figure 2b.

## Vibrational Spectroscopy of Polyvinyl Chloride

We use infrared (IR) absorption spectroscopy[S1] to gain general insights into the compositions of the commercial PVC films used in these experiments. These spectra, for PVC1 and PVC2, are shown in Figure S5. Although an in-depth analysis of these spectra is well outside the scope of this work, here we briefly consider selected absorption peaks. Some peaks can be unambiguously assigned to the PVC polymer, such as the three peaks at ≈ 608 cm$^{-1}$, ≈ 635 cm$^{-1}$, and ≈ 698 cm$^{-1}$, which are associated with C–Cl stretches, and the H–C–Cl out-of-plane angular deformation at ≈ 1254 cm$^{-1}$. We note that the peak at ≈ 610 cm$^{-1}$ is the most distinctive of PVC, even in the presence of plasticizers that may have peaks overlapping other bands.[S2] By contrast, the prominent carbonyl (C=O) stretching peak, at ≈ 1725 cm$^{-1}$, likely originates from one or more additives blended into the film, since this moiety is not present in the structure of PVC but is found in many common additives (e.g., phthalate or adipate plasticizers). In the spectrum of PVC2, this assignment is corroborated by the doublet at ≈ 1580 cm$^{-1}$ and ≈ 1600 cm$^{-1}$, which is characteristic of di(2-ethylhexyl)phthalate,[S2] and by the presence of aromatic C–H stretching vibrations at 3024 cm$^{-1}$ and 3061 cm$^{-1}$. Other peaks in the PVC2 spectrum (but not in the PVC1 spectrum) that suggest phthalate plasticizers are observed at ≈ 743 cm$^{-1}$, ≈ 1040 cm$^{-1}$ (associated with C–H bending in the aromatic ring),[S3,S4] and ≈ 1254 cm$^{-1}$, which are consistent with previous observations.[S2] By contrast, the spectrum of PVC1 has a prominent doublet at ≈ 1143 cm$^{-1}$ and ≈ 1170 cm$^{-1}$, which likely indicates a different additive or plasticizer. Many other peaks in the IR spectra have more ambiguous origins, such as the methylene (CH$_2$) and methyl (CH$_3$) stretching modes between 2800 cm$^{-1}$ and 3000 cm$^{-1}$, and various other C–H modes in the fingerprint region, since these excitations are expected from both the PVC and plasticizer film constituents. Overall, the spectra differences may indicate that PVC2 contains more aromatic-derived additives (e.g., phthalate plasticizers) than PVC1. Finally, we note that these IR spectra are acquired using an attenuated total internal reflectance (ATR) geometry. Accordingly, they are subject to ATR-related peak shifts and distortions relative to spectra measured in other configurations (e.g., transmission or photothermal induced resonance spectra), though these differences are typically small and (aside from their presence) are not significant for this analysis.

These results confirm that the film composition (e.g., different blends of PVC and plasticizer) also affects the PVC working temperatures, as demonstrated previously.[S5] The plasticizer and its mixing ratio with PVC plays a critical role in determining the working temperature range (pick-up and release temperatures) of the PVC films. In principle, researchers determined to optimize film properties, or to assess the behaviors of films not commercially available, could test films with a wide range of composition and thicknesses. For most researchers, however, commercial sources will likely be sufficient to assemble a wide variety of 2D material heterostructures and devices.

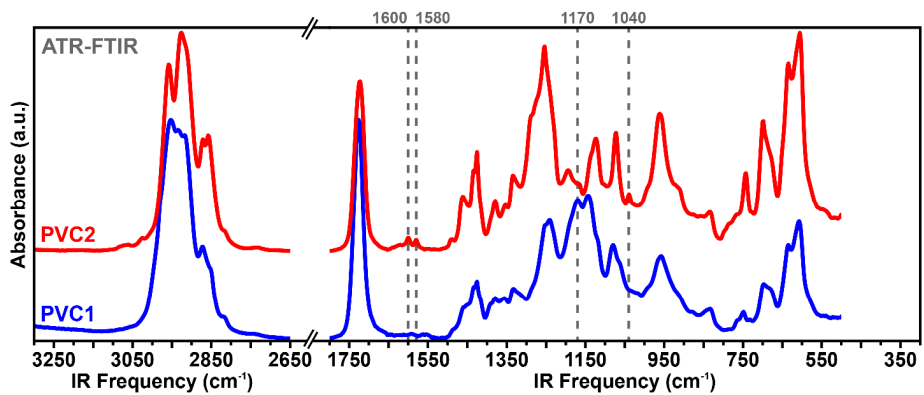

**Figure S5:** Fourier transform infrared (FTIR) absorption spectroscopy of polyvinyl chloride (PVC) films. (a) Comparisons of absorption spectra measured for each of the two commercially available PVC films (PVC1 and PVC2) used in these experiments. As expected, these spectra exhibit peaks characteristic of not only PVC but also one or more additives (e.g., plasticizers) present in the films. Despite their many similarities, differences in the spectra measured for PVC1 and PVC2, some of which are indicated with dashed vertical lines, suggest different film compositions (e.g., in the types and relative proportions of plasticizers). Spectra are measured in an attenuated total internal reflectance (ATR) geometry, shown without correction, with normalized intensities, and vertically offset for clarity.

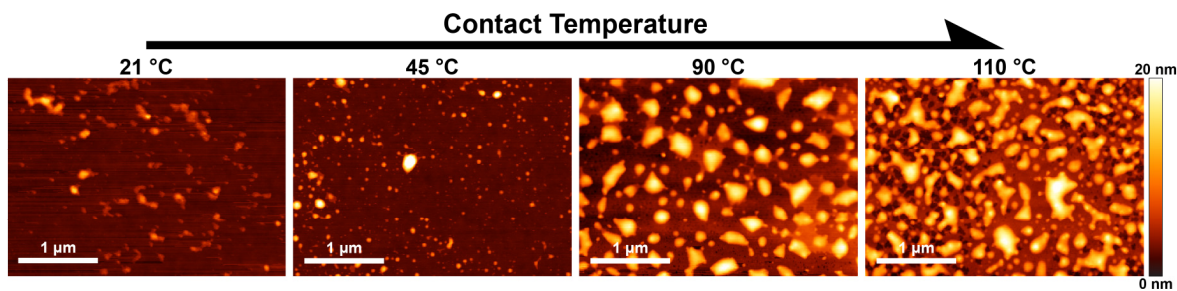

**Figure S6:** Atomic force topographs of $SiO_2$/Si substrates contacted at the indicated temperatures (21 °C: Room temperature; 45 °C: PVC1 pickup temperature; 90 °C: PVC1 release temperature; 110 °C: PVC1 stamp conditioning temperature) by unconditioned polymer stamps with bare PVC1 working surfaces (i.e., no two-dimensional materials). Contaminants, transferred from the stamp to the contacted surface, appear as irregularly shaped protrusions on the otherwise essentially featureless surface. More contaminants are transferred from the stamp at higher contact temperatures, with typical features heights ≲ 15 nm at 110 °C. Similar trends are also observed for PVC2 working surfaces (not shown), with higher stamp/surface contact temperatures appearing more contaminated by PVC-derived residue. To enable easier side-by-side comparisons, the height (color) scale in all images is set to a common range (20 nm), with the plane of the substrate approximately equalized in each panel. Regions outside of this height range are shown saturated at the limits displayed.

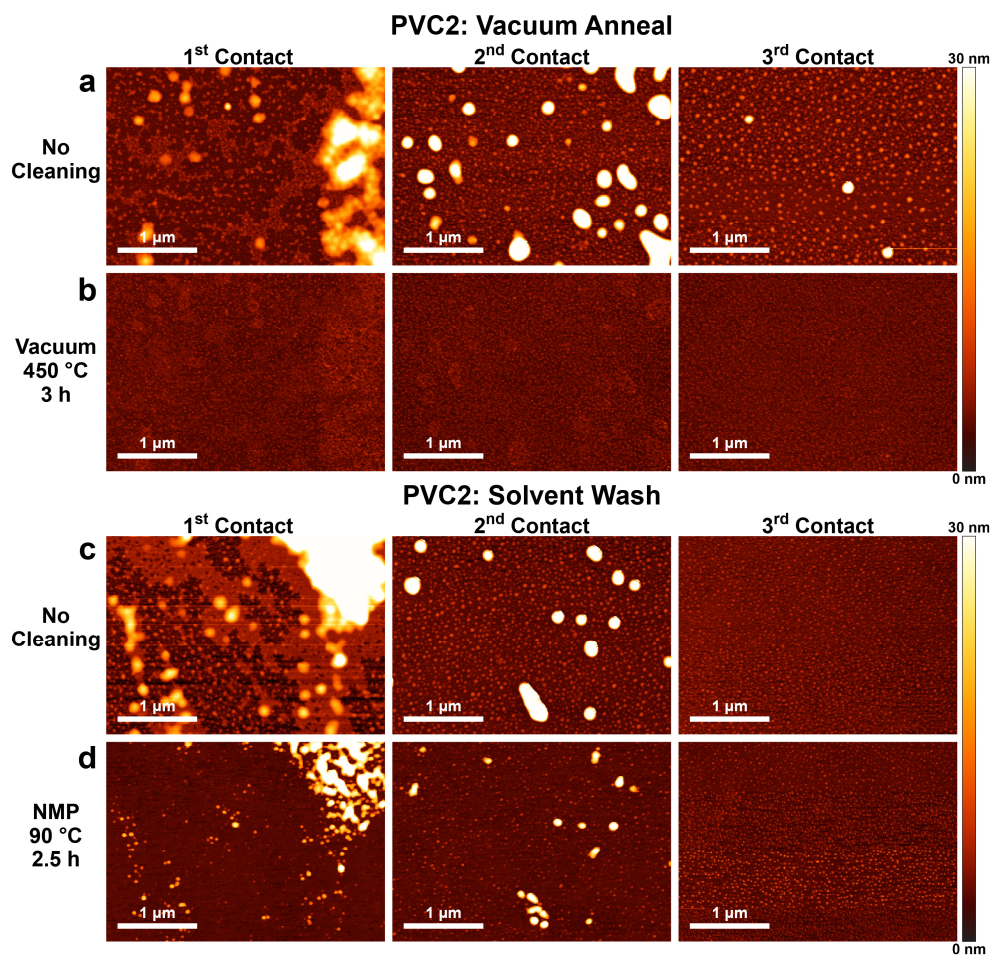

**Figure S7:** Atomic force topographs of $SiO_2$/Si substrates contacted by polymer stamps with bare PVC2 working surfaces (i.e., no two-dimensional materials). Images in the same row compare different regions of the substrate contacted sequentially by the same stamp, indicated with ordinal numbers: 1st, 2nd, 3rd, at 160 °C. (a, c) After removing the stamps, the contacted regions of the substrates are nearly covered with irregularly shaped contaminants transferred from the stamps, with some aggregates protruding several tens of nanometers. Contaminants are most prominent in the first-contact regions, containing the largest-area, most-protruding surface features, with progressively less contamination in the second- and third-contact regions. (b, d) These contaminants are substantially removed after cleaning with either the vacuum annealing or solvent washing protocols described in the text. (b) The same regions shown in (a) after annealing in vacuum at ≈ 450 °C for about 3 h. (d) The same regions shown in (c) after washing in N-methyl-2-pyrrolidone (NMP) at ≈ 90 °C for about 2.5 h. To enable easier side-by-side comparisons, the height (color) scale in all images is set to a common range (30 nm), with the plane of the substrate approximately equalized in each panel. Regions outside of this height range are shown saturated at the limits displayed. We note that an imaging artifact is evident in the image of the third-contact, solved-washed regions (i.e., bottom-right-most panel), likely corresponding to a temporary change in the quality of the probe tip and not reflective of significant differences in sample topography within this region. See also Figures 5 and S8.

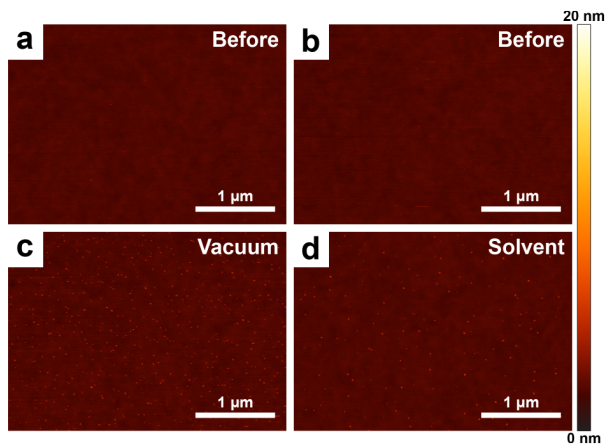

**Figure S8:** Atomic force topographs of bare regions of $SiO_2$/Si substrates subjected to different cleaning processes. These regions are *never contacted directly* by polyvinyl chloride (PVC) stamps but are located on the same substrates shown in Figure 5, contacted by PVC1 stamps and later cleaned. (a, b) Bare regions of the substrates, prior to any stamping, with no significant topographic features. (c) The same region shown in (a) after annealing in vacuum at ≈ 350 °C for about 3 h to clean PVC residue at other locations on the same substrate (see Figure 5a). (b) The same region shown in (b) after washing in N-methyl-2-pyrrolidone at ≈ 90 °C for about 3 h to clean PVC residue at other locations on the same substrate (see Figure 5b). After cleaning, under careful inspection, both regions show very small (≈ 1 nm in height) protrusions, resembling the features on the cleaned, PVC-contacted surfaces in Figure 5. In the case of the vacuum-annealed substrate, we hypothesize that these features result from aggregation of Au atoms, sourced from metal alignment marks not shown in the scanned field of view, which diffuse across the $SiO_2$ surface during annealing. We note that these features are also observed far (≈ millimeters) from the gold alignment marks on the same substrate, though with a lower areal density than near the alignment marks as shown here. In the case of the solvent-cleaned substrate, the remaining particles might result from (re)deposition of material from the liquid phase during washing. In both cases, however, we consider the surfaces clean from significant PVC residue and expect further optimization of the washing technique (or elimination of mobile metal species from the surface) to resolve the features observed here. To enable easier side-by-side comparisons, the height (color) scale in all images is set to a common range (20 nm), with the plane of the substrate approximately equalized in each panel.

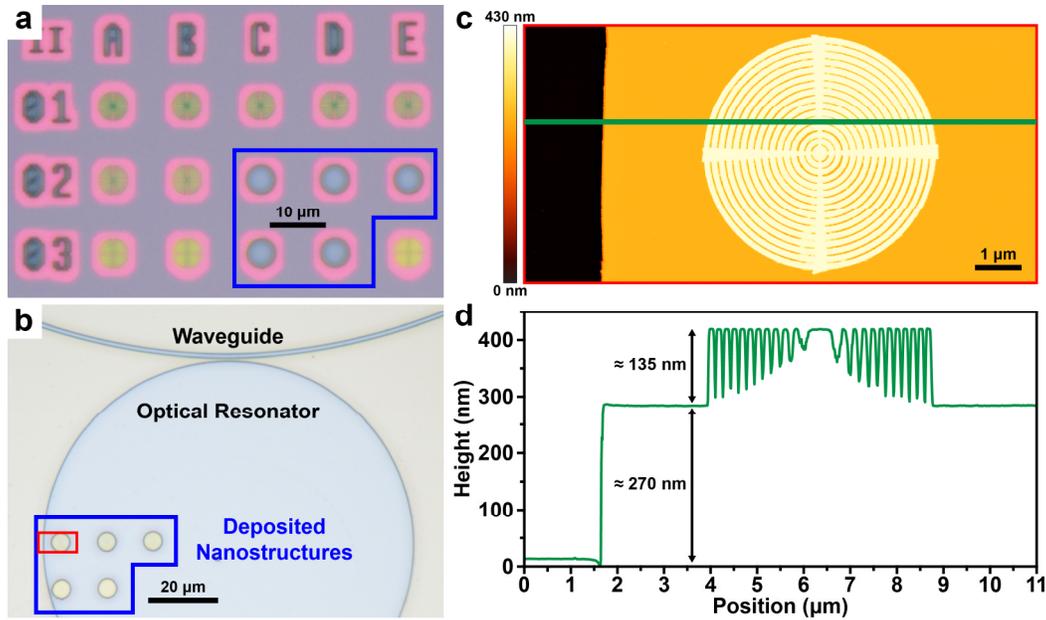

**Figure S9:** Controlled pick up and deposition of $Al_{0.4}Ga_{0.6}As$ nanostructures to create a heterogeneously integrated photonic device. (a) Optical microscope image of an array of $Al_{0.4}Ga_{0.6}As$ nanostructures. Although too small to resolve their detailed features, optically, these "bullseye" structures are composed of concentric rings (each $\approx$ 80 nm wide) surrounding a central disk (diameter $\approx$ 285 nm), with a total diameter of $\approx$ 5 µm. Importantly, these structures are partially released from the underlying substrate and suspended over a thin, air-filled cavity by 4 spoked tethers connecting to the sounding $Al_{0.4}Ga_{0.6}As$ film. As described in the main text, a PVC1 stamp picked up a group of 5 $Al_{0.4}Ga_{0.6}As$ nanostructures from the region outlined in blue in (a). (b) Optical microscope image showing the 5 transferred $Al_{0.4}Ga_{0.6}As$ nanostructures, outlined in blue, deposited near the edge of a $Si_3N_4$ optical resonator and waveguide fabricated on a $SiO_2/Si$ substrate. (c) Atomic force topograph of one of the deposited nanostructures, outlined in red in (b), showing its fine spatial features, after vacuum annealing at $\approx$ 400 °C for 10 h. (d) Topographic profile, measured along the green path indicated in (c), spanning the edge of the optical resonator ($\approx$ 270 nm thick) and $Al_{0.4}Ga_{0.6}As$ nanostructure ($\approx$ 135 nm thick).

**Table S1:** Summary of materials tested for pick up with polyvinyl chloride (PVC) stamps.

| Material | Material Thickness | | Ease of pick up[b] |
| --- | --- | --- | --- |
| | Monolayer | Bulk[a] | |
| hBN | **Yes** | **Yes** | Very Easy |
| Graphene | No[c] | **Yes** | May require repeated attempts |
| $MoS_2$ | **Yes** | **Yes** | May require repeated attempts |
| $WSe_2$ | **Yes** | **Yes** | May require repeated attempts |
| $MoTe_2$ | **Yes** | **Yes** | May require repeated attempts |
| $NbSe_2$ | Not tested | No | N/A |
| $TaS_2$ | Not tested | No | N/A |
| $Fe_5GeTe_2$ | Not tested | **Yes** | Easy |
| $Fe_3GeTe_2$ | Not tested | No | N/A |
| $CuInP_2S_6$ | Not tested | **Yes** | Easy |
| $In_2Se_3$ | Not tested | **Yes** | Easy |
| CrSBr | Not tested | **Yes** | May require repeated attempts |
| $Al_xGa_{1-x}As$[d] | N/A | **Yes** | Easy |

[a]Here, "bulk" loosely refers to flakes of two-dimensional materials composed of more than about 3 or 4 stacked layers, with thicknesses more than $\approx$ 2 nm.

[b]Subjective judgment by the authors on the relative ease and reliability that flakes of the specified materials transfer from $SiO_2$/Si substrates to PVC1 surfaces at temperatures of $\approx$ 45 °C, using the process describe in the main text. Pick-up characteristics may change under other conditions.

[c]Can pick up monolayer graphene by using a hBN handle (to pick up monolayer edge first).

[d]Can pick up bulk ($\approx$ 135 nm thick) GaAs/$Al_{0.4}Ga_{0.6}As$/GaAs slabs, released from the underlying substrate, as described in main text.

# Supplementary References